\documentclass[twocolumn,aps,superscriptaddress,prx]{revtex4-2}

\usepackage{bm}%
\usepackage{float}
\expandafter\ifx\csname package@font\endcsname\relax\else
 \expandafter\expandafter
 \expandafter\usepackage
 \expandafter\expandafter
 \expandafter{\csname package@font\endcsname}%
\fi

\usepackage{comment}
\usepackage{array}
\usepackage{amsmath,amssymb}
\usepackage{MnSymbol}
\usepackage{tikz-feynman,contour} 
\usetikzlibrary{shapes,arrows,positioning,automata,backgrounds,calc,er,patterns}
\usepackage{feynmf}
\tikzfeynmanset{compat=1.0.0} 
 
\usepackage{graphicx}
\usepackage{mathrsfs}
\usepackage{bm}
\usepackage{mathtools}
\usepackage{epstopdf}
\usepackage{hyperref}

\hypersetup{%
   pdfpagemode=None, 
   pdfstartpage=1,
   pdfmenubar=true,
   pdftoolbar=true,
   colorlinks = true,
   linkcolor=blue,
   citecolor=blue,
   urlcolor=blue,
   bookmarksopen=false
 }
 \usepackage{amsmath}
\usepackage{amsfonts}
\usepackage{mathtools}
\usepackage{graphicx}
\usepackage{fixme}
\usepackage{color}
\usepackage{cancel}
\usepackage{bm}

\usepackage{epstopdf}
\usepackage{float}
\usepackage{amsmath}
\usepackage{color}
\usepackage{comment}
\usepackage{dsfont}
\usepackage{braket}
\usepackage[normalem]{ulem}

\newcommand\identity{1\kern-0.25em\text{l}}

\usepackage{feynmf}
\usepackage{tikz-feynman}
\usetikzlibrary{shapes,arrows,positioning,automata,backgrounds,calc,er,patterns}
\tikzfeynmanset{compat=1.0.0}

\newcommand\bwt         {\begin{widetext}}
\newcommand\ewt         {\end{widetext}}

\def\bi{{\bf i}}
\def\bj{{\bf j}}

\def\bk{{\bf k}}
\def\bp{{\bf p}}
\def\bq{{\bf q}}
\def\bk{{\bf k}}

\def\br{{\bf r}}
\def\bd{{\bf d}}

\definecolor{airforceblue}{rgb}{0.36, 0.54, 0.66}
\definecolor{amber}{rgb}{1.0, 0.75, 0.0}
\definecolor{applegreen}{rgb}{0.55, 0.71, 0.0}
\definecolor{alizarin}{rgb}{0.82, 0.1, 0.26}

\newcommand{\jens}[1]{{\color{purple}#1}}

\begin{document}

\title{Spin-Charge Bound States and Emerging Fermions in a Quantum Spin Liquid}
\author{Jens H.\ Nyhegn}
 \affiliation{Department of Physics and Astronomy, Aarhus University, Ny Munkegade, DK-8000 Aarhus C, Denmark}
 \author{Kristian Knakkergaard Nielsen}
 \affiliation{Niels Bohr Institute, University of Copenhagen, Jagtvej 128, DK-2200 Copenhagen, Denmark}
 \affiliation{Max-Planck Institute for Quantum Optics, Hans-Kopfermann-Str. 1, D-85748 Garching, Germany}
 \author{Leon Balents}
 \affiliation{Kavli Institute for Theoretical Physics, University of California, Santa Barbara, California 93106-4030, USA}
 \affiliation{French American Center for Theoretical Science, CNRS, KITP, Santa Barbara, California 93106-4030, USA}
\affiliation{Canadian Institute for Advanced Research, Toronto, Ontario, Canada}
 \author{Georg M.\ Bruun}
\affiliation{Department of Physics and Astronomy, Aarhus University, Ny Munkegade, DK-8000 Aarhus C, Denmark}

\begin{abstract}
The complex interplay between charge and spin dynamics lies at the heart of  strongly correlated quantum materials, and it is a fundamental topic in basic research with far reaching technological perspectives. We explore in this paper the dynamics of holes in a single band extended $t$-$J$ model where the background spins form a $\mathbb{Z}_{2}$ quantum spin liquid. Using a field theory approach based on a parton construction, we show that while the electrons for most momenta fractionalize into uncorrelated charge carrying holons and spin carrying spinons as generally expected for a quantum spin liquid, the spinon-holon scattering cross-section diverges for certain momenta signalling strong correlations. By deriving an effective low-energy Hamiltonian describing this dynamics, we demonstrate that these divergencies are due to the formation of long lived spinon-holon bound states. Since the wave function of these bound states are localized over a few lattice sites, they correspond to well-defined fermions with the same charge and spin as the underlying electrons. 
We then show that quantum gas microscopy with atoms in optical lattices provides an excellent platform for verifying and probing the internal spatial structure of these emerging fermions.
The  fermions will furthermore show up as clear quasiparticle peaks in angle-resolved photoemission spectroscopy  with an intensity determined by their internal structure. For a nonzero hole concentration, the fermions form hole pockets with qualitatively the same location, shape, and intensity variation in the Brillouin zone as the so-called Fermi arcs observed in the pseudogap phase. Such agreement is remarkable since the Fermi arcs arise from the delicate interplay between the symmetry of the quantum spin liquid  and the internal structure of 
the emerging fermions in a minimal single band model with no extra degrees of freedom added.
Our results, therefore, provide a microscopic mechanism for the  conjectured fractionalized Fermi liquid  and  open up new pathways for exploring the pseudogap  phase and high temperature superconductivity as arising from a  quantum spin liquid.
\end{abstract}

\maketitle

\section{Introduction}
The reduced dimensionality of two-dimensional (2D) materials renders their properties highly sensitive to quantum mechanical effects. This gives rise to exotic phenomena such as long-range quantum entanglement, quasiparticles with nontrivial statistics, and spin-charge separation~\cite{Keimer2017}. A prominent example is quantum spin liquids (QSLs) consisting of a macroscopic superposition of spin singlet configurations covering the lattice, an idea which, since it was introduced more than 50 years ago, continues to attract significant attention with many fundamental and open questions~\cite{anderson1973,savary2017,zhou2017,broholm2020}. A QSL was recently observed with the use of Rydberg atoms in an optical tweezer array~\cite{semeghini2021}, and there is promising progress toward realizing a QSL with atoms in optical lattices~\cite{Yamamoto_2020,Yang2021,sun2023,Lebrat2024,Prichard2024}. 
Diverse solid-state materials with kagome and triangular lattices show signs of QSL physics~\cite{scheie2024proximate,bag2024evidence,khuntia2020gapless}.  QSLs are also predicted to exist in the rapidly growing class of 2D materials consisting of twisted atomically thin layers~\cite{Pan2020,Zang2021,Morales2022, song2025}. 
A prevailing paradigm regarding   QSLs is that their low-energy excitations have lost any resemblance to the underlying electrons, 
which instead have ”fractionalized” into largely uncorrelated spin and charge degrees of freedom~\cite{PhysRevB.29.3035, lee2006}. It has, however, been suggested 
 that a new kind of fractionalized Fermi liquid
(FL*) can exist in such QSLs when doped away from half-filling. This may, moreover, be connected to the puzzling pseudogap phase 
observed in the cuprates~\cite{senthil2003,yang2006,kaul2007,qi2010,moon2011,punk2012, vojta2012,punk2015,Zhang2020,bonetti2024, PhysRevB.73.174501},
following an early idea that high-temperature superconductivity arises from a QSL~\cite{anderson1987,Kivelson1987,Kotliar1988,Balents1998,PWAnderson2004}.

\begin{figure}[t!]
	\begin{center}
	\includegraphics[width=\columnwidth]{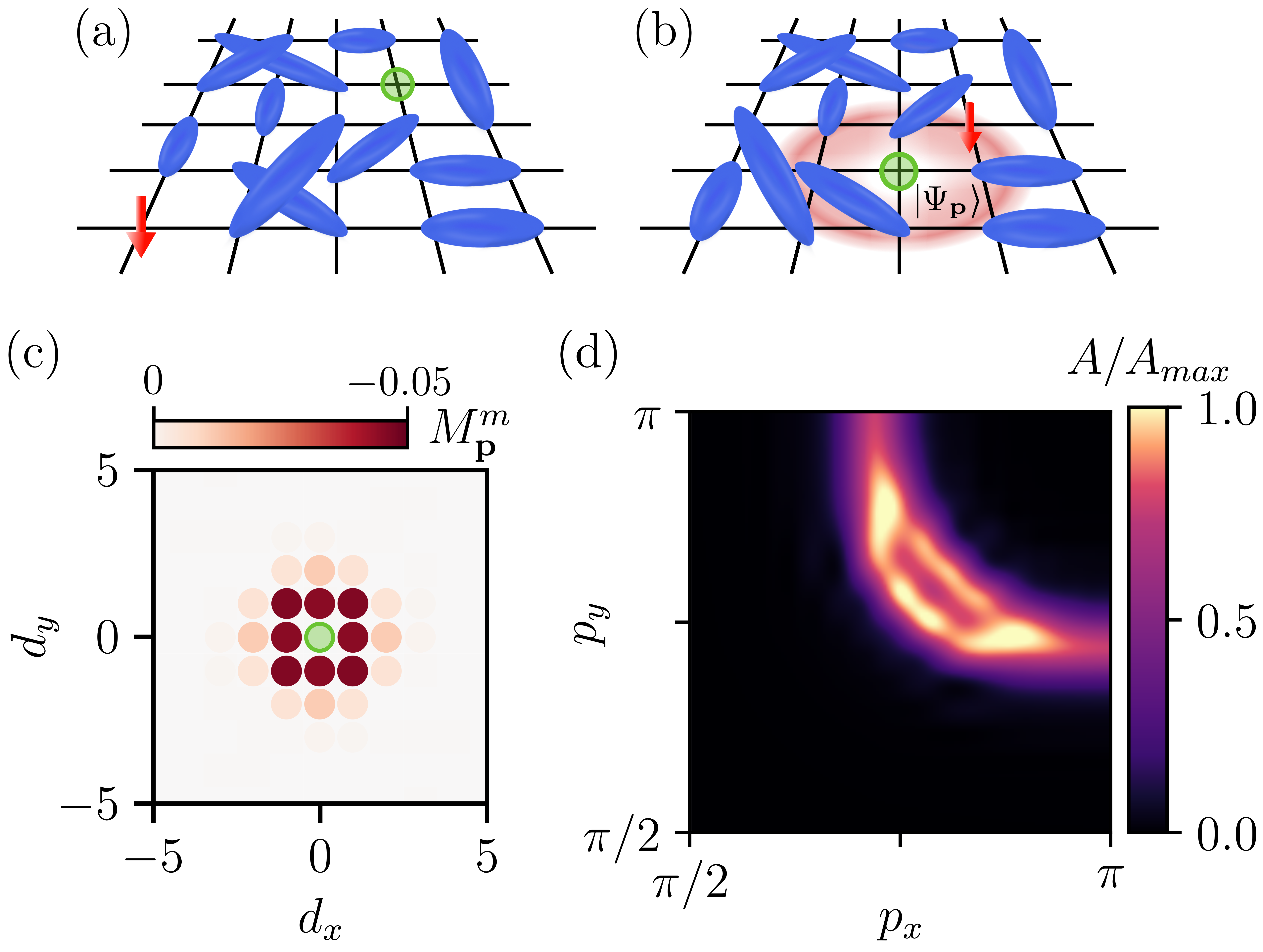}
	\end{center}
	\caption{(a) A physical spin $\uparrow$ hole fractionalizes into an unbound spinon (arrow) and holon (green circle) pair in a spin liquid formed by 
    spin singlets (blue ovals). (b) For some momenta, however, the holon and spinon bind to form a new fermionic quasiparticle with well-defined spin and charge, and a 
     spatial structure determined by the relative spinon-holon wave function $|\Psi_{\mathbf p}\rangle$ extending over a few lattice sites as indicated by the red coloring. (c) Mimicking the spatial structure of the bound state, the hole-spin correlation function $M^{n}_{\bp}(\bd)$ effectively describes the likelihood of finding the unpaired spin at a given lattice with the holon (green circle) at the center. 
    (d) The spectral response in the Brillouin zone of these fermions at their Fermi energy as measured by angle-resolved photoemission spectroscopy for a nonzero hole doping $x=0.01$,
    where they form a fractionalized Fermi liquid. This clearly shows a hole pocket around $\bp=(3\pi/4,3\pi/4)$  with a dimly lit back side due to the interplay between the internal symmetry of the fermions and the quantum spin liquid singlets. 
    There are identical hole pockets in the other corners of the Brillouin zone. The shape and spectral weight of these pockets are similar to those of the Fermi arcs observed in the pseudogap phase of
    the cuprates. 
    }
	\label{fig.FrontPage}
\end{figure}

Inspired by this, we explore in this paper the properties of holes in a $\mathbb{Z}_{2}$ quantum spin liquid background described by a single-band, extended $t$-$J$ model on a square lattice close to half-filling. We develop a field-theoretic framework based on a parton construction, from which we derive an effective low-energy Hamiltonian describing the hole dynamics in terms of charge-carrying holons 
interacting with spin-carrying spinons. While the spinons and holons are essentially uncorrelated for most center-of-mass momenta (COM)
in agreement with the prevailing paradigm for QSLs,  
we show that they can form long-lived bound states in certain regions in the Brillouin zone (BZ). These bound states have a relative wave function localized over a few lattice sites and, therefore, describe emerging fermions with a well-defined energy, charge, and spin. We then demonstrate that the internal spatial structure
of these fermions can be probed with the use of well-established quantum gas microscopy techniques in optical lattices, and that they show up as sharp peaks in ARPES experiments.  
Remarkably, for a nonzero concentration of holes they 
form  pockets in the BZ giving rise to angle-resolved photoemission spectroscopy (ARPES) spectra 
with qualitatively the same position and shape as the intriguing Fermi arcs observed in the pseudogap phase. 
This shape arises from the subtle interplay between the symmetries of the spin liquid and the internal structure 
of the fermions emerging from the $t$-$J$ model with no fitting parameters or extra degrees of freedom added. 
As such, our theory provides a microscopic basis for the conjectured existence of a FL* state in the pseudogap phase of the cuprates. Our main results are shown in Fig.~\ref{fig.FrontPage}. 

This paper is organized as follows. Section \ref{sec.model} introduces the extended $t$-$J$ model together with the parton description of holes in the spin liquid. In Sec.\ \ref{sec:FieldTheory}, we discuss our field-theoretic approach to calculate the hole Green's function, and in Sec.\ \ref{sec.effSch} we derive an effective Schr\"odinger equation giving a nonperturbative description of the holon-spinon 
correlations. The emergence and properties of spinon-holon bound states are discussed in Sec.\ \ref{sec.BoundState}, and we explore in Sec. \ref{sec.observables} how these bound states 
can be probed with quantum gas microscopy  in optical lattices  as well as with ARPES 
in condensed-matter systems.
We end with conclusions and an outlook in Sec. \ref{sec.DiscOut}.  



´

\section{Model} \label{sec.model}
We consider spin $\sigma=\uparrow,\downarrow$ fermions on a two-dimensional square lattice with nearest-neighbor $t_1$ and next-nearest-neighbor $t_2$ hopping and on-site repulsion $U$. This is described by a single-band Hubbard model, 
which in  the limit of strong repulsion $t_i/U\ll 1$ close to half-filling (one electron per site) reduces to the extended $t$-$J$ model $\hat H=\hat H_t+\hat H_J$. Here, 
\begin{equation}
	\hat{H}_t=  -t_{1}\sum_{\langle \bi, \bj\rangle,\sigma}( \tilde{c}^{\dagger}_{\bi,\sigma} \tilde{c}^{\vphantom\dagger}_{\bj,\sigma} + \text{h.c.}) 
	 -t_{2}\sum_{\llangle \bi, \bj\rrangle,\sigma}( \tilde{c}^{\dagger}_{\bi,\sigma} \tilde{c}^{\vphantom\dagger}_{\bj,\sigma} +\text{h.c.}),
	\label{eq.Ht}
\end{equation}
is the kinetic energy, with $\langle \bi, \bj\rangle$ nearest-neighbor and $\llangle \bi, \bj\rrangle$ next-nearest-neighbor  lattice sites,  and
 $\tilde{c}^{\dagger}_{\bi,\sigma} = \hat{c}^{\dagger}_{\bi,\sigma}(1-\hat{n}_{\bi})$, where $\hat c_{\bi\sigma}$ removes a fermion with spin $\sigma$ at site $\bi$ so that  $\hat{n}_{\bi}=\sum_\sigma\hat{c}^{\dagger}_{\bi,\sigma}\hat{c}_{\bi,\sigma}$ is the density operator. The antiferromagnetic Heisenberg term is
\begin{equation}
	\hat{H}_J = \ J_{1}\sum_{\langle\bi, \bj\rangle}  \hat{\mathbf S }_{\bi}\cdot \hat{\mathbf S }_{\bj}  +J_{2}\sum_{\llangle \bi, \bj \rrangle} \hat{\mathbf S }_{\bi}\cdot \hat{\mathbf S }_{\bj} ,
	\label{eq.HJ}
\end{equation}
where $\hat{ {\bf S} }_{\bf i} = \frac{1}{2}\sum_{\sigma,\sigma'} \hat{ c }^\dagger_{{\bf i},\sigma}\boldsymbol{\sigma}_{\sigma\sigma'}\hat{ c }_{{\bf i},\sigma'}$, where $\boldsymbol{\sigma}=(\sigma_x,\sigma_y,\sigma_z)$ are spin operators and  $J_{i} = 4t^{2}_{i}/U$. 
We have  in Eq.\ \eqref{eq.HJ}  left out a density-density interaction $-\sum_{\braket{\bi,\bj}}\hat{n}_{\bi}\hat{n}_{\bj}/4$, since it becomes a 
constant for a single hole injected into a half-filled background. 




At half-filling the model reduces to the $J_{1}$-$J_{2}$ Heisenberg model $\hat H_J$ whose ground state is predicted to be a $\mathbb{Z}_{2}$ QSL in a finite region of the phase diagram for spin couplings around $J_{2}/J_{1}=0.5$~\cite{hu2013, wang2018, ferrari2018, ferrari2020, yu2018, PhysRevB.38.9335}. Our goal is to study the hole dynamics in this QSL phase, and we use a parton construction where the electron operator is decomposed as
$\hat{ c }_{\bi,\sigma} =\hat{b}^{\dagger}_{\bi} \hat{f}_{\bi,\sigma}$.   Here, 
$\hat{f}_{\bi,\sigma}$ removes a fermionic spinon, which carries spin but no charge, and $\hat{b}_{\bi}$ removes a bosonic holon carrying charge but no spin~\cite{lee2006}. With the use of this construction, the Heisenberg term $\hat H_J$ becomes quartic in the 
spinon operators and different spin liquids can then be described using different mean-field theories as a starting 
point~\cite{wen2010}.
By comparison with results from variational Monte Carlo  calculations~\cite{hu2013, yu2018}, it was found that 
\begin{align}
	\hat{H}^\text{mf}_{J} =  \sum_{\bk, \sigma} \epsilon_{\bk} \hat{ f }^{\dagger}_{\bk,\sigma} \hat{ f }_{\bk,\sigma}  + \sum_{\bk} \left( \Delta_{\bk} \hat{f}^{\dagger}_{\bk,\uparrow} \hat{f}^{\dagger}_{-\bk,\downarrow} + \text{h.c.}\right) 
	\label{Eq.spinon} 
\end{align}
provides a mean-field description of the $\mathbb{Z}_2$ QSL as well as its low-energy 
excitations~\cite{ferrari2019}. Here, 
\begin{align}
\epsilon_{\bk} &= \epsilon(\cos k_{x} + \cos k_{y}) \nonumber \\
\Delta_{\bk} &=\Delta_{1}(\cos k_{x}-\cos k_{y}) + \Delta_{2}\sin{(2k_{x})}\sin{(2k_{y})}
\label{eq.spinon_def}
\end{align} 
are the dispersion and $d_{xy}$-wave pairing fields, describing the singlet formation in the QSL, $\bk$ is a crystal momentum in the BZ, and the lattice constant is taken to be unity. We have $\Delta_{1} = 1.8 \epsilon$ and 
$\Delta_{2} = 1.1 \epsilon$   for $J_{2}/J_{1}=0.5$~\cite{yu2018}, and  from Ref.~\cite{ferrari2019} we estimate $\epsilon \approx J_{1}/3$. In the following, we fix $J_{1}/J_{2}=0.5$, such that the only free parameter is the interaction strength $t_{1}/J_1$ controlling the competition between the charge motion and spin ordering. We set $t_{2}/t_{1}=-\sqrt{J_{2}/J_{1}}$ to describe the case of hole doping; $t_{2}/t_{1}= + \sqrt{J_{2}/J_{1}}$ would instead describe electron doping \cite{jiang2021}.

Diagonalization of Eq.~\eqref{Eq.spinon} yields
\begin{align}
	\hat{H}^\text{mf}_{J} =&   \sum_{\bk, \sigma} \omega^{s}_{\bk} \hat{\gamma}^{\dagger}_{\bk\sigma} \hat{\gamma}_{\bk\sigma}, 
	\label{Eq.spinon_diag}
\end{align}
with $\hat\gamma_{\bk\sigma}$ removing a spinon with momentum $\bk$, spin $\sigma$, and
energy $\omega^{s}_{\bk}=\sqrt{\epsilon_{\bk}^2+\Delta_{\bk}^2}$. 
We have $\hat f_{\bk\uparrow}=u_{\bk}\hat \gamma_{\bk\uparrow}-v_{\bk}\hat\gamma^\dagger_{-\bk\downarrow}$
and $\hat f_{\bk\downarrow}=u_{\bk}\hat \gamma_{\bk\downarrow}+v_{\bk}\hat\gamma^\dagger_{-\bk\uparrow}$
with the coherence factors $u_{\bk} = \sqrt{( 1 + \epsilon_{\bk}/\omega^{s}_{\bk})/2}$ and $v_{\bk} = \text{sgn}(\Delta_{\bk}) \sqrt{( 1 -\epsilon_{\bk}/\omega^{s}_{\bk})/2}$.

When holes are introduced, their motion can destroy the singlets of the QSL. This 
is described by the hopping term, which  reads 
\begin{align}
	\hat{H}_{t} = &\sum_{ \bk }\omega^{h}_{\bp} \hat{b}^{\dagger}_{\bp} \hat{b}_{\bp}+  
	 \sum_{ \bp, \bk,\bq, \sigma } h_{\bp,\bk,\bq} \hat{b}^{\dagger}_{\bp-\bq} \hat{b}_{\bp} \hat{\gamma}^{\dagger}_{\bk+\bq, \sigma} \hat{\gamma}_{\bk, \sigma} \nonumber \\ 
	+ & \sum_{ \bp, \bk,\bq }  g_{\bp,\bk,\bq} \left( \hat{b}^{\dagger}_{\bp-\bq} \hat{b}_{\bp} \hat{\gamma}^{\dagger}_{\bk+\bq, \downarrow} \hat{\gamma}^{\dagger}_{\bk, \uparrow} + \text{h.c.} \right)
	 \label{Eq.H_inter} 
\end{align}
in terms of partons~\cite{nyhegn2025}.  Here 
\begin{align}
	g_{\bp,\bk,\bq} &=\Lambda_{\bp-\bk}u_{\bk} v_{\bk-\bq} +  \Lambda_{\bp+\bk-\bq}  v_{\bk} u_{\bk-\bq}  \nonumber  \\ 
	h_{\bp,\bk,\bq} &=\Lambda_{\bp-\bk-\bq} u_{\bk+\bq} u_{\bk} -\Lambda_{\bp+\bk} v_{\bk+\bq} v_{\bk}   
	\label{Eq.vertex}
\end{align}
are the vertex functions for the holon scattering on a spinon and emitting/absorbing two spinons, and 
\begin{align}
	\omega^{h}_{\bp} &= 2 \sum_{\bk} \Lambda_{\bp+\bk}  v_{\bk}^{2} 
\end{align}
is the bare holon dispersion with 
\begin{align}
	\Lambda_\bk = -\frac{2}{N} \Big(& t_{1}\left[ \cos{k_{x}} + \cos{k_{y}} \right] \nonumber \\ 
	+ &t_{2}\left[ \cos{(k_{x}+k_{y})} + \cos{(k_{x}-k_{y})} \right] \Big).
    \label{Gamma}
\end{align}

We focus in this paper on the hole spectrum, which can be measured with ARPES as discussed in  Sec.~\ref{SpectralSec}. The spectrum related to inserting an extra electron is separated by a large energy gap $U$ since it involves at least one lattice site with double occupancy. 
\section{Field theory}\label{sec:FieldTheory}
\begin{figure}[t!]
	\begin{center}
    \includegraphics[width=0.99\columnwidth]{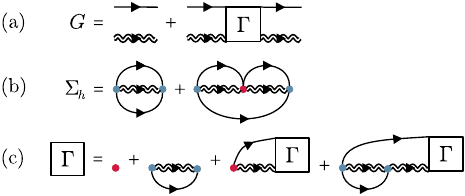}
    \end{center}
	\caption{(a) The Green's function $G(\bp,\omega)$ of a physical hole is a two-body Green's function of a spinon and a holon. Solid lines are the spinon propagator and double wavy lines are the  dressed holon propagator Eq.~\eqref{eq.holonprop}. (b) Holon self-energy $\Sigma_h(\bp,\omega)$ giving the dressing of holons by spinons. (c) Spinon-holon scattering matrix $\Gamma(\bp,\bq_1,\bq_2\omega)$, with red and blue balls the scattering $h$ and $g$ vertices, respectively, in Eq.~\eqref{Eq.vertex}. 
    }
    \label{eq.GreenScat}
\end{figure}

We now describe our field-theoretic approach for analyzing the dynamics of holes in the QSL. 
Removal of a spin $\uparrow$ electron from the QSL at half-filling is described by  the  retarded hole Green's function 
\begin{gather}
	G(\bp,\tau) = -i\theta(\tau)\langle \{ \hat{c}^{\dagger}_{\bp,\uparrow}(\tau), \hat{c}_{\bp,\uparrow}(0)\}\rangle\nonumber\\ = -\frac{i\theta(\tau)}{N}\!\sum_{\bq_{1},\bq_{2}} v_{\bq_{1}}v_{\bq_{2}} 
	\left\langle \hat{\gamma}_{\bq_{2},\downarrow}(\tau) \hat{b}_{\bp - \bq_{2}}(\tau) \hat{b}^{\dagger}_{\bp - \bq_{1}}  \hat{\gamma}^{\dagger}_{\bq_{1}, \downarrow}  \right\rangle,
	\label{eq.prop_hole_rot}
\end{gather}
where $\hat A(\tau)=\exp(i\hat H\tau)A\exp(-i\hat H\tau)$ is an operator at time $\tau$ in the Heisenberg picture. 
As is apparent from Eq.~\eqref{eq.prop_hole_rot}, the one-body Green's function for a physical hole is a \emph{two-body} 
spinon-holon Green's function in the parton picture, since the removal of a spin $\uparrow$ electron creates 
a holon and a spin $\downarrow$ spinon. Calculation of this two-body Green's function, 
shown diagrammatically in Fig.~\ref{eq.GreenScat}(a),
is challenging for strong interaction $t_1/J_1>1$ due to the spinon-holon interactions 
given in Eq.~\eqref{Eq.H_inter}. These interactions enter in two qualitatively different ways. 

First, while the spinons are unaffected by the presence of a single hole so that their propagator is given by $1/(\omega-\omega_\bk^s)$ in frequency space, the holons can be strongly affected by 
the dressing with spinons in the QSL.  We use a self-consistent diagrammatic approach to 
   include this dressing for the holon propagator, which in frequency space reads 
\begin{equation}
	G_h(\bp,\omega)=\frac1{\omega-\omega_{\bp}^h-\Sigma_h(\bp,\omega)}.
	\label{eq.holonprop}
\end{equation}
This  amounts to 
resumming a class of noncrossing  diagrams for the holon self-energy $\Sigma_h(\bp,\omega)$ to infinite order as shown 
in Fig.~\ref{eq.GreenScat}(b) and described in detail in Ref.~\cite{nyhegn2025,FootnoteNotation}. Our approach is analogous to the self-consistent Born approximation  for hole motion in an antiferromagnet~\cite{kane1989}, which 
has proven to be very accurate when compared with  experimental and exact numerical results~\cite{martinez1991,diamantis2021,nielsen2022}.
Inclusion of this dressing of the holon propagator but assuming that it moves independently of the 
the spinon corresponds to the first diagram in Fig.~\ref{eq.GreenScat}(a). Such uncorrelated 
motion is equivalent to the hole fractionalizing into an  unbound spinon-holon pair.
It gives rise to a broad spectral response of the hole as  
  was analyzed in detail in Ref.~\cite{nyhegn2025}, and this
 absence of a well-defined quasiparticle peak has been 
proposed  as a characteristic feature of  QSLs~\cite{laeuchli2004, kadow2022, kadow2024}.
A second way of interaction enters when one is calculating the hole Green's function between the holon and the spinon that are simultaneously created when the electron is removed. 
Such interactions are described by the holon-spinon scattering matrix $\Gamma$ entering the second diagram in
Fig.~\ref{eq.GreenScat}(a).
Given the diagrams we use to calculate the holon Green's
function shown in Fig.~\ref{eq.GreenScat}(b), we must include the diagrams generated by the Bethe-Salpeter equation 
shown in Fig.~\ref{eq.GreenScat}(c) to 
obtain a conserving approximation for the two-body holon-spinon Green's function~\cite{baym1962}. Doing this to infinite order is, however,
very challenging, and in practice, we instead numerically iterate the Bethe-Salpeter equation order by order,
which is then used to calculate 
the hole Green's function. 
As detailed in  App.~\ref{app.FullCalc}, we find that convergence is quickly obtained after a few iterations for most momenta and
that the spinon-holon interactions have little effect on the hole Green's function, which is well-described by the 
uncorrelated first diagram in Fig.~\ref{eq.GreenScat}(a). This  gives a broad spectral response coming from the fractionalization of the electron as  
expected for a spin liquid. For momenta in the vicinity of $\bk=(\pi,\pi)$, we however cannot obtain 
convergence with the hole spectral function changing significantly from one iteration to the next. 
We now show that this behavior is due to a singularity arising from the presence of a bound state that cannot be described by finite-order perturbation theory.

\section{Effective Schr\"odinger equation} \label{sec.effSch}
We now   derive an effective Schr\"odinger equation for the spinon-holon pair, which automatically 
includes the spinon-holon scattering events  in $\Gamma$ to infinite order in a much simpler way than solving the full 
Bethe-Salpeter equation.
 
To do this, we first notice that diagrams involving the spinon pair creation/annihilation
vertex $g$ (blue vertices in Fig.~\ref{eq.GreenScat}(c)) turn out to give small contributions and do not capture the divergences; see App.~\ref{app.FullCalc} for details. 
It is, therefore, sufficient to include the dominant ladder diagrams involving the vertex $h$ (diagrams with red vertices in Fig.~\ref{eq.GreenScat}(c)) to identify any bound states, which describe repeated spinon-holon scatterings to infinite order. 
This gives the Bethe-Salpeter equation 
\begin{align}
        \Gamma(\bp,\bk,\bk',\omega) &= h_{\bp,\bk,\bk'} + \sum_{\bq}h_{\bp,\bk,\bq} G(\bp -\bq,\omega - \omega^{s}_{\bk+\bq})\nonumber \\
     &  \times\Gamma(\bp-\bq,\bk+\bq,\bk'-\bq,\omega).
     \label{BetheSalpeter}
\end{align}

We then use a pole expansion of the holon Green's function, writing  
\begin{align}
	G_h(\bp,\omega) \simeq \frac{Z^{h}_{\bp}}{\omega - \tilde\omega^{h}_{\bp}+i\gamma_{\mathbf p}^{h}},
	\label{eq.poleExpand}
\end{align}
where $\tilde\omega^{h}_{\bp}$ is the energy of a dressed holon with momentum $\bp$, determined by solving $\tilde\omega^{h}_{\bp} = \omega_{\bp}^h+\text{Re}\Sigma_h(\bp,\tilde\omega^{h}_{\bp})$
numerically, $Z^{h}_{\bp}$ is its residue, and $\gamma_{\mathbf p}^h= -Z^{h}_{\bp} \text{Im}\Sigma_h(\bp,\tilde\omega^{h}_{\bp})$ is half its the decay rate~\cite{nyhegn2025}, which can be nonzero due to spinon-holon interactions. 

Finally, we define an effective interaction vertex $\tilde{h}_{\bp,\bk,\bk'} = (Z^{h}_{\bp})^{1/2}h_{\bp,\bk,\bk'}(Z^{h}_{\bp - \bk'})^{1/2}$ and scattering matrix $\tilde{\Gamma}(\bp,\bk,\bk',\omega) = (Z^{h}_{\bp })^{1/2} \Gamma(\bp,\bk,\bk',\omega)(Z^{h}_{\bp - \bk'})^{1/2}$ so that  Eq.~\eqref{BetheSalpeter} can be written as
\begin{align}
    &\tilde{\Gamma}(\bp,\bk,\bk',\omega) = \tilde{h}_{\bp,\bk,\bk'} \nonumber \\
     &+ \sum_{\bq} \frac{\tilde{h}_{\bp,\bk,\bq}}{\omega \!-\! \omega^{s}_{\bk+\bq} \!-\! \tilde\omega^{h}_{\bp-\bq} \!+\! i\gamma_{\bp - \bq}^{h}} \tilde{\Gamma}(\bp\!-\!\bq,\bk\!+\!\bq,\bk'\!-\!\bq,\omega).
    \label{eq.scat.renorm}
\end{align} 
With Eq.~\eqref{eq.scat.renorm}, we have reduced the complicated Bethe-Salpeter  equation to a much simpler Lippmann-Schwinger equation describing the two-body scattering of a spinon-holon pair. This in turn is equivalent to a Schr\"odinger equation $\hat{H}_\text{eff} \ket{\Psi} = E\ket{\Psi}$ for the relative wave function $\ket{\Psi}$ of the spinon-holon pair 
with the effective Hamiltonian~\cite{sakurai2021}
\begin{align}
	\hat{H}_\text{eff} =& \sum_{\bk}(\tilde\omega^{h}_{\bk}-i\gamma^{h}_{\mathbf k})\tilde{b}^{\dagger}_{\bk}\tilde{b}_{\bk} + \sum_{\bk}\omega^{s}_{\bk}\hat{\gamma}^{\dagger}_{\bk}\hat{\gamma}_{\bk} \nonumber \\
    &+ \sum_{\bk,\bk'}\tilde{h}_{\bp-\bk,\bk,\bk'-\bk}  \tilde{b}^{\dagger}_{\bp-\bk'} \hat{\gamma}^{\dagger}_{\bk', \sigma} \tilde{b}_{\bp-\bk} \hat{\gamma}_{\bk, \sigma}.
\label{Hamiltonian}
\end{align} 
Here, $\tilde{b}^\dagger_{\bk}$ creates a dressed holon quasiparticle  with momentum $\bk$, energy $\tilde\omega^{h}_{\bk}$,
and decay rate $\gamma^{h}_{\mathbf k}$. Equation \eqref{Hamiltonian} provides a general and physically intuitive description of hole dynamics in the QSL in terms of dressed holons interacting with spinons, which is much easier to analyze than the full field theory.
Nontrivial many-body effects of the background QSL are included in several ways. First, the dispersion $\omega^{s}_{\bk}$ of the spinons is determined by the background QSL and the dispersion $\tilde \omega^\text{h}_{\bk}$ of the holons includes the dressing by spinons. Second, the spinon-holon interaction is renormalized by the holon residues in analogy with the microscopic foundation of Fermi liquid theory~\cite{negele1988quantum,PhysRevX.8.031042}. It also depends on all momenta, reflecting the lack of Galilean invariance, making it nonlocal, which is typical for effective Schr\"odinger equations for many-body systems. Finally, the dressed holons can be damped due to their interactions with the spinons, and Eq.\ \eqref{Hamiltonian} in this sense describes an open system. 
It was shown in Ref.~\cite{nyhegn2025} that the dressed holons have a vanishing decay rate around the momentum $\bp= (\pi,\pi)$,
whereas they become strongly damped and eventually disappear in a region around  $\bp= (0,0)$.
We determine the energy of these strongly damped holons from their spectral function peak, but since they have a small 
or vanishing residue they essentially do not interact with any other states and thus have little influence on the dynamics.



Since we have used a number of approximations to arrive at Eq.~\eqref{Hamiltonian},  a natural question  is the accuracy of this approach. 
Here it should be  noted that we have used a similar mapping of a Bethe-Salpeter equation to an effective Schr\"odinger equation to analyze the bound states 
between two mobile impurities in a Bose-Einstein condensate. This turned out to be remarkably accurate when compared with 
Monte-Carlo calculations even for strong interactions~\cite{camacho-guardian2018a,massignan2025polaronsatomicgasestwodimensional}.

\section{Bound states and emerging fermions} \label{sec.BoundState}
We now use the effective Hamiltonian given by Eq.~\eqref{Hamiltonian} to 
analyse the two-body spinon-holon problem created when an electron is removed from the QSL. In
particular, we show that the divergences described in Sec.~\ref{sec:FieldTheory}  when iterating the 
second diagram in Fig.~\ref{eq.GreenScat}(a) are due to the 
presence of spinon-holon bound states. 

To do this, we solve the Schr\"odinger equation 
\begin{equation}
\hat H_\text{eff}\ket{\Psi^{n}_{\bp}}=E^{n}_{\bp}\ket{\Psi^{n}_{\bp}}
\label{Schrodinger}
\end{equation}
with 
\begin{equation}
    \ket{\Psi^{n}_{\bp}} = \sum_{\bk}\phi^{n}_{\bp,\bk}\tilde{b}^{\dagger}_{\bp-\bk}\hat{\gamma}^{\dagger}_{\bk,\downarrow}\ket{\text{QSL}}
	\label{eq.Eig}
\end{equation}
the $n$th two-body eigenstate with  COM  momentum $\bp$. Here, $\phi^{n}_{\bp,\bk}$ is the relative spinon-holon wave function and $\ket{\text{QSL}}$ is the QSL ground state at half-filling with 
$\hat{b}_{\bp}\ket{\text{QSL}}=\hat{\gamma}_{\bk,\sigma}\ket{\text{QSL}}=0$. Solving Eq.~\eqref{Schrodinger} numerically is much easier than solving the full Bethe-Salpeter equation.

\begin{figure}[t!]
	\begin{center}
	\includegraphics[width=0.48\textwidth]{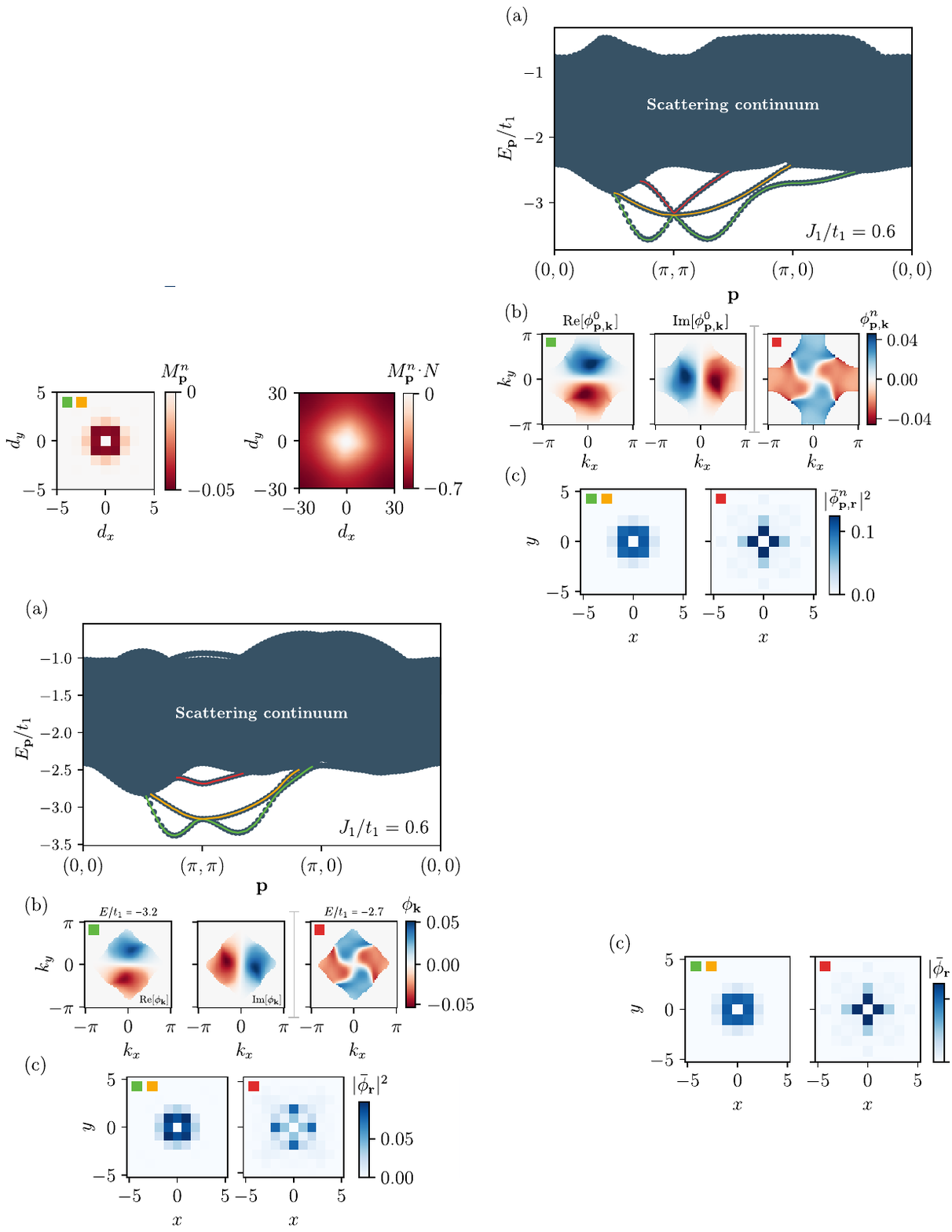}
	\end{center}
	\caption{(a) Spinon-holon energy spectrum obtained from  Eqs.~\eqref{Schrodinger}-\eqref{eq.Eig} for a  $64\times64$ 
    lattice along 
    straight paths in the Brillouin zone. The red, yellow, and green bands are bound states. 
    (b) Bound state wave functions for $\bp=(\pi,\pi)$ in momentum space. The two leftmost panels illustrate the real and imaginary parts of one of the degenerate $p$-wave states. For the other $p$-wave state, the imaginary part has changed sign, while the real part is identical. (c) Plots of the gauge-averaged relative wave function in real space, with the left panel showing the $p$-wave states. This shows that the spin and charge of the quasiparticles are located within a few lattice sites.  }
	\label{Fig.EnergySpectrum}
\end{figure}

In Fig.\ \ref{Fig.EnergySpectrum}(a), we plot the energy spectrum of the spinon-holon states $\ket{\Psi^{n}_{\bp}}$ as a function of their COM along straight lines in the BZ  for the interaction strength $J_{1}/t_{1} = 0.6$. The 
spectrum is determined by taking the real part of the complex energy obtained by solving Eq. \eqref{Schrodinger} numerically. We see that the solutions fall into two distinct  classes. First, there is a continuum coming from delocalized scattering states dominated by a single term in the superposition in Eq.~\eqref{eq.Eig}. These states are the unbound  spinon-holon pairs, i.e.\ the fractionalized fermions expected for a QSL. Notably, we  also find three discrete energy bands well below the continuum as highlighted with green, yellow, and red lines. These states, moreover, have a small decay rate on the order of $\sim 10^{-2}$ times the energy gap to the continuum, making them well-defined \emph{bound states}. The reason for their long lifetime is that the bound states are essentially uncoupled from the strongly damped but also weakly interacting holons; see  App.~\ref{app.NH} for details. 


We plot in  Fig.\ \ref{Fig.EnergySpectrum}(b) the wave functions $\phi^{n}_{\bp,\bk}$ for bound states at the COM momentum ${\bp}=(\pi,\pi)$. At this momentum, the green band is degenerate with the yellow band whereas the red band has a slightly higher energy. Figure \ref{Fig.EnergySpectrum}(c) shows the wave function squared in real space obtained by Fourier transform. 
These plots clearly show that the relative wave function is spread out in momentum space and localized in real space, 
reflecting that the spinon and holon are bound together. 
 We also see that the wave functions of the two lowest degenerate states have a $p$-wave symmetry, whereas the highest energy bound state wave function has 
 $d$-wave symmetry. Note that the wave functions in momentum space do not have the mirror symmetries of the lattice due to our 
  specific choice of phase (gauge) of the mean-field $\Delta_\bk$ in Eq.~\eqref{eq.spinon_def} describing the singlets of the QSL. We have averaged over this 
 gauge for the real-space plot in  Fig.\ \ref{Fig.EnergySpectrum}(c), thereby recovering these symmetries. This accounts for the different symmetry-broken states realized in repeated experiments. It should be noted that at half-filling, these symmetries are naturally present 
  due to an SU$(2)$ gauge  \cite{wen2002}: see Appendix \ref{app.gaugeAVG} for details.
 
From this, we conclude that the QSL for certain COM momenta supports bound spinon-holon states with energies below the usual scattering continuum, a long lifetime, and an  internal spatial structure extending over a few lattice sites.
These states, therefore, have a well-defined energy, spin, and charge   and they  correspond to new electronlike fermions.
We emphasize that these fermions emerge from our theory based on the single-band Hubbard model 
in the strong-coupling limit with no need to add further degrees of freedom, and that they have no obvious  
smooth connection to the electronic quasiparticles present at weak interactions. 
This is a main result of the present paper and it provides a microscopic mechanism for the conjecture that the pseudogap phase of the 
 cuprates consists of a so-called fractionalized Fermi liquid (FL*) existing on top of a QSL~\cite{senthil2003,yang2006,kaul2007,qi2010,moon2011,punk2012, vojta2012,punk2015,bonetti2024,Zhang2020}. 
 We explore the FL* phase existing for a finite concentration of holes further in Sec.~\ref{SpectralSec}. 
In App.~\ref{app.LS}, we show by tuning the pairing potential that the formation of these quasiparticles is stabilized by shorter singlets in the QSL.

\section{Observables} \label{sec.observables}
 We now turn to how these emerging fermions and the FL* can be observed in condensed matter as well as optical lattice systems. 
 Since the fermions are bound states of spinons and holons with a very long lifetime from the low-energy part of the spectrum, we ignore the holon decay in the following. 
 In App.~\ref{app.NH} it is shown that this is a very good approximation.

\subsection{Spatial structure}
Using quantum microscopy with atoms in optical lattices, one can take pictures of 
many-body systems with single-site resoluton~\cite{doi:10.1126/science.aal3837}. 
In particular, the spin correlations have been measured in the neighborhood of mobile holes (or particles) added to a 
Fermi Hubbard system at half-filling~\cite{koepsell2019a,ji2021a,koepsell2021a,Lebrat2024,Prichard2024}.

In the parton construction, a vacant site means it is occupied by a holon, and we, therefore, analyze the correlation function 
 \begin{align}
    &M^{n}_{\bp}(\bd) = \frac{\langle\Psi^{n}_{\bp}|\hat{S}^{z}_{\br+\bd} \hat{b}^{\dagger}_{\br}\hat{b}_{\br}|\Psi^{n}_{\bp}\rangle}{\langle\Psi^{n}_{\bp}|\hat{b}^{\dagger}_{\br}\hat{b}_{\br}|\Psi^{n}_{\bp}\rangle} \nonumber  \\ 
    &= -\frac{1}{2N}\left(\left|\sum_{\bk}\phi^{n}_{\bp,\bk}u_{\bk}e^{-i\bk\cdot \bd}\right|^{2}
    + \left|\sum_{\bk}\phi^{n}_{\bp,\bk}v_{\bk}e^{-i\bk\cdot \bd}\right|^{2}\right)
	\label{eq.M}
\end{align}
for the spin $\hat{S}^{z}_{\br+\bd}=(\hat{f}^{\dagger}_{\br+\bd,\uparrow}\hat{f}_{\br+\bd,\uparrow} - \hat{f}^{\dagger}_{\br+\bd,\downarrow}\hat{f}_{\br+\bd,\downarrow})/2$
observed at $\br+\bd$ given that the holon is observed at $\br$, assuming that the system is in the quantum state $|\Psi^{n}_{\bp}\rangle$. The normalization gives
 $\langle\Psi^{n}_{\bp}|\hat{b}^{\dagger}_{\br}\hat{b}_{\br}|\Psi^{n}_{\bp}\rangle=1/N$  since the holon is uniformly distributed
in the lattice. We also have $\sum_{\bd} M^{n}_{\bp}(\bd)=-1/2$, reflecting that a single spin-up electron has been removed from the half-filled QSL.

\begin{figure}[t!]
	\begin{center}
	\includegraphics[width=0.49\textwidth]{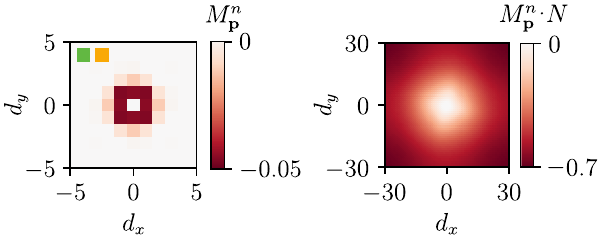}
	\end{center}
	\caption{Spin-hole correlation function $M^{n}_{\bp}(\bd)$ given by Eq.~\eqref{eq.M} for $\bp =(\pi,\pi)$, $J_{1}/t_{1}=0.6$, and a $64\times 64$ lattice. The left panel shows the $p$-wave bound state, and the right panel shows a 
   delocalized state in the scattering continuum. Note that the scale in the right panel is multiplied by $N$. }
	\label{Fig.MPi}
\end{figure}

The  left panel in Fig.\ \ref{Fig.MPi} shows the spin-hole correlation function $M^{n}_{\bp}(\bd)$ for the lowest bound state $|\Psi^{n}_{\bp}\rangle$ with momentum $\bp=(\pi,\pi)$. We have again performed a gauge averaging to simulate the shot-to-shot 
variations in an experiment; see Appendix \ref{app.gaugeAVG}. This clearly shows that the net spin (magnetization) 
is spatially localized in the 
vicinity of the hole, reflecting that the spinon and holon have formed a fermionic quasiparticle with well-defined charge and spin localized over a few lattice sites. 
The symmetry of the magnetization around the hole is directly inherited from the relative spinon-holon wave function 
shown in Fig.~\ref{Fig.EnergySpectrum}(c). In contrast, the right panel in Fig.~\ref{Fig.MPi} shows $M^{n}_{\bp}(\bd)$ for one of the scattering states. Here, the magnetization is of order $1/N$ on each lattice site and is distributed over the whole lattice even when the hole is observed at a give lattice 
site. This explicitly shows the fractionalization of the fermions into unbound charge (holon) and spin (spinon) excitations  characteristic of QSLs~\cite{lee2006}.


Note that we omitted the dressing of the holons by spinons as described by $\Sigma_h$ when deriving the second line in Eq.~\eqref{eq.M}. This dressing cloud would also frustrate the singlet environment surrounding the holon, but this dressing is SU$(2)$ symmetric, so $\langle\hat{S}^{z}_{\br}\rangle=0$ everywhere for an isolated dressed holon, reflecting that it carries charge but no spin.  This makes it possible for $M^{n}_{\bp}(\bd)$ to distinguish between bound spinon-holon pairs and uncorrelated quasiparticles.


These results, therefore, show that optical lattices are a promising platform for observing the internal spatial structure of the emerging fermions in the FL*, 
provided that the relevant parameter and temperature regimes of the Hubbard model can be realized. In fact, 
the spin correlation function  $M^{n}_{\bp}(\bd)$ is  \emph{simpler} than those already measured around holes in a 
Fermi-Hubbard antiferromagnet, which involve at least two spin operators~\cite{koepsell2019a,ji2021a,koepsell2021a,Lebrat2024,Prichard2024}. 

\subsection{Spectral properties}\label{SpectralSec}
ARPES is a very successful technique for probing the spectral properties   
of condensed-matter systems. It has, in particular been used to observe  the characteristic  "Fermi arcs" in 
the pseudogap phase of unconventional superconductors~\cite{yang2011,he2011,doi:10.1126/science.aaw8850}, 
whose microscopic origin remains an open and important question~\cite{bonetti2024}. 
The spectral properties of magnons in optical lattices have furthermore  recently been measured by radio-frequency  
spectroscopy~\cite{prichard2025observationmagnonpolaronsfermihubbardmodel}, and this technique 
 can  be used to probe the quasiparticles formed by mobile holes in a Fermi-Hubbard system~\cite{nielsen2025dualspectroscopyquantumsimulated}.

\begin{figure}[t!]
	\begin{center}
	\includegraphics[width=0.48\textwidth]{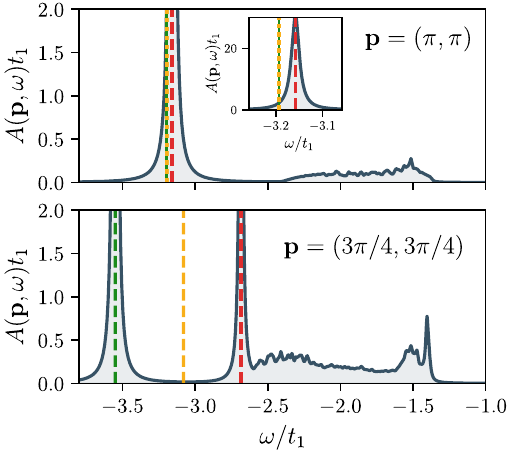}
	\end{center}
	\caption{Hole spectral function given by Eq.~\eqref{eq.spec} for two different momenta. The vertical dashed lines 
    show the energy of the bound states with the same color code as in Fig.~\ref{Fig.EnergySpectrum}. The inset shows an enlarged section around the peak. }
	\label{Fig.SpecWeight}
\end{figure}

We, therefore,  analyze the spectral properties of the FL* as probed by these techniques. The contribution to the hole spectral function $A(\bp,\omega)=-2\text{Im}G(\bp,\omega)$ coming from the uncorrelated (fractionalized) spinon-holon pairs was analyzed in Ref.~\cite{nyhegn2025} by evaluation of the first  diagram in 
Fig.~\ref{eq.GreenScat}(a). Here, we focus on  spinon-holon bound states as described by the second diagram in Fig.~\ref{eq.GreenScat}(a). 
With the use the Lehman representation and Eq.~\eqref{eq.Eig}, this can be written as 
\begin{align}
    A(\bp,\omega) = \frac{2}{N}\sum_n\frac{\eta}{\left(\omega-E^{n}_{\bp}\right)^{2} + \eta^{2}}\left|\sum_{\bk}v_{\bk}\phi^{n}_{\bp,\bk}\right|^{2},
    \label{eq.spec}
\end{align}
where we have added the generic small imaginary part,  $\eta = 0.01t_{1}$, also used when the analytical continuation for the holon's Green's function is performed. In the following plots, Figs. \ref{Fig.SpecWeight} and \ref{Fig.FS2}, we have again performed a gauge averaging of the spectral function.

Consider first the spectral function for the high-symmetry momentum $\bp = (\pi,\pi)$ shown in Fig.~\ref{Fig.SpecWeight}(a). 
  The vertical dashed lines indicate the energies $E_{\bp}^{n}$ of the bound states shown in Fig.~\ref{Fig.EnergySpectrum}. 
We see that the highest bound state gives rise to a clear quasiparticle  peak in the spectrum, whereas 
 the two lowest states 
are \emph{not} visible, i.e.\ they are "dark".  This is due to the sum $\sum_{\bk}\phi^{n}_{\bp,\bk}v_{\bk}$  in Eq.~\eqref{eq.spec}, giving the overlap between the spinon-holon wave function and the state $\hat{c}_{\bp,\uparrow}\ket{\text{QSL}} = \sum_{\bk}v_{\bk}\hat{b}^{\dagger}_{\bp-\bk}\hat{\gamma}^{\dagger}_{\bk,\downarrow}\ket{\text{QSL}}/N$ created by removal of an electron from the QSL. The state created by removal of an electron, $\hat{c}_{\bp,\uparrow}\ket{\text{QSL}}$, 
inherits the $d_{xy}$-wave symmetry of the QSL given by Eq.~\eqref{eq.spinon_def}, which makes 
 its overlap with the two lowest states at $\bp = (\pi,\pi)$  vanish because they have $p$-wave symmetry. The overlap, on the other hand, is nonzero for the third state since it has $d$-wave symmetry. In Fig. \ref{Fig.SpecWeight}(b), the spectral function is plotted for $\bp = (3\pi/4,3\pi/4)$ where the bound states have a lower symmetry. They 
are consequently all bright in the hole spectral function, although the spectral weight of the middle (yellow) energy state is  very small because the contribution from the $d_{xy}$-wave pairing channel is small. 

To further illustrate how the interplay between the internal structure of the fermions and the symmetry of the singlet in the QSL 
show up in the spectral properties, we plot in Fig.~\ref{Fig.FS2}(a) the spectral function $A(\bp,\omega)$ along straight lines in the BZ. 
This shows that the yellow (middle) band in Fig. \ref{Fig.EnergySpectrum} has a very low spectral weight, making it almost dark. 
We also see that the spectral weight of the lowest band decreases when we move from the minimum around $\bp=(3\pi/4,3\pi/4)$ toward the high-symmetry point $\bp=(\pi,\pi)$, where the interaction wave function becomes $p$-wave symmetric, making it dark as discussed above.

These results demonstrate that the emerging fermions can be detected in ARPES as  peaks in the hole spectral function 
with a spectral weight determined by how their internal spatial symmetry relates to the symmetry of the QSL state.
 


\subsection{fractionalized Fermi liquid and  Fermi arcs}
In analogy with the weakly interacting Feshbach dimers formed by strongly bound atoms~\cite{RevModPhys.80.1215}, we 
expect the emerging fermions to be weakly interacting when their binding energy is large so that their spatial size is a few lattice sites. For  a nonzero  hole concentration $x$, they will then form a so-called fractional Fermi liquid. In the following, we focus on the low-doping regime, where spin-liquids have been found to remain stable, and examine how this manifests itself in ARPES experiments \cite{PhysRevB.86.085145, PhysRevLett.119.067002}.

\begin{figure}[t!]
	\begin{center}
	\includegraphics[width=\columnwidth]{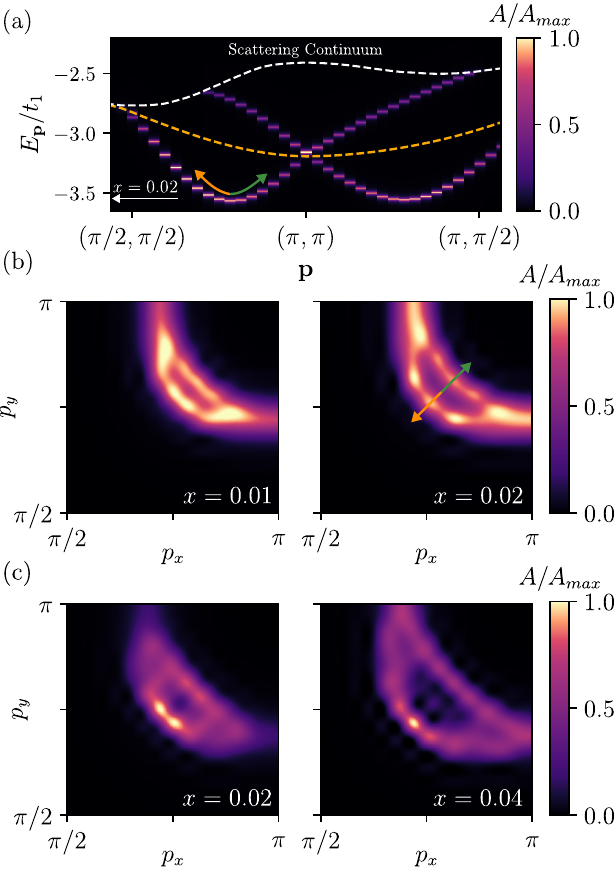}
	\end{center}
	\caption{(a) Spectral function $A(\bp,\omega)$ along straight paths in the Brillouin zone for $J_1/t_1=0.6$. The dashed yellow line indicates the position of the yellow (middle) band in Fig. \ref{Fig.EnergySpectrum}. The white arrow indicates the Fermi energy for the filling fraction $x=0.02$ used in (b). Spectral function at the Fermi surface of the FL* for the indicated hole doping $x$ and for (b) $J_1/t_1=0.6$ and (c) $J_1/t_1=0.3$. These clearly display hole pockets with dimly lit back sides. There are three identical hole pockets in the other corners of the Brillouin zone. The orange and green arrows in (b) indicate the same directions in the Brillouin zone as in (a). }
	\label{Fig.FS2}
\end{figure}

Figures ~\ref{Fig.FS2}(b) and ~\ref{Fig.FS2}(c) show the spectral function in the BZ for different Fermi energies $\varepsilon_F$ of the FL* 
corresponding to the hole-filling fractions $x=0.01$, $x=0.02$, and $x=0.04$, and for different values of $J_1/t_1$; see also Fig.~\ref{fig.FrontPage}.
That is, $\varepsilon_F$ is the highest energy of the occupied  spinon-holon bound states for given filling 
fraction $x$. We  see   that the holes form anisotropic  pockets located 
at the corners of the BZ with a low-intensity back side. Comparing Fig. \ref{Fig.FS2} (b) and (c), we see that the arc features are robust with respect to doping and interaction strength, whereas their exact shape and size depend 
quite sensitively on these parameters. Again, this suppression of intensity is a result of these bound states predominately having $p$-wave symmetry, giving a 
suppressed overlap with the $d$-wave singlets. The green and orange arrows in Fig.~\ref{Fig.FS2}(b) indicate the paths in the BZ  in 
 Fig.~\ref{Fig.FS2}(a), where the Fermi level for $x=0.02$ is also shown.   
Note that the area of the hole pockets in Figs. \ref{Fig.FS2}(b) and \ref{Fig.FS2}(c) is given by their concentration $x$, which is quite different from the area $1+x$ of weakly interacting holes smoothly connected to the noninteracting holes.  Luttinger's theorem~\cite{LuttingerWard1960,Luttinger1960} is, therefore, broken by the FL*, showing that it is distinct from the Fermi liquid present at weak interactions~\cite{senthil2003,kaul2007}.

Remarkably, the  hole pockets in Fig.~\ref{Fig.FS2} are qualitatively similar to the 
Fermi arcs measured with ARPES in the pseudogap phase of the cuprates~\cite{yang2011,he2011,doi:10.1126/science.aaw8850,he2014, fujita2014}, with respect to both their 
shape and their anisotropic spectral weight. 
Recovering  these features  is highly nontrivial since  our microscopic theory is based on a single-band 
$t$-$J$ model from which the FL*  emerges for strong interactions 
as a Fermi liquid of bound spinon-holon states with no extra ingredients required. 
The predicted Fermi arcs are a result of the delicate interplay between the energies of these fermions as well as 
their internal spatial structure. 
Our results, therefore, provide a microscopic mechanism for the conjecture that the  pseudogap phase is  a FL*  coexisting with a QSL 
background, which has been used to explain quantum oscillation and transport measurements, as well as the observed Fermi arcs~\cite{lee2006, qi2010,vojta2012, yang2006, punk2012, punk2015, bonetti2024, moon2011, kaul2007,bonetti2024}. 
To our knowledge, most calculations supporting this conjecture have so far been based on 
phenomenological descriptions or models where the fermions effectively are added on top of the background 
QSL on the basis of physical arguments.

\section{Discussion and Outlook} \label{sec.DiscOut}
Using a field-theoretic approach, we explored the dynamics of holes in a background $\mathbb{Z}_{2}$ QSL as described by a single-band, extended $t$-$J$ model close to half-filling. The spinon-holon scattering was shown to diverge for certain COM, and we then derived an effective low-energy Hamiltonian to demonstrate that this divergence is due to spinon-holon bound states. These bound states correspond to the existence of well-defined fermions with the same charge and spin as the underlying electrons, 
which is unexpected from the general paradigm of uncorrelated spin and charge degrees of freedom in QSLs. 
A nonzero concentration of holes is then shown to lead to hole pockets whose ARPES spectra exhibit the qualitative features of the  Fermi arcs observed in the cuprate pseudogap phase with respect to both the shape and the position in the Brillouin zone. 

Since the shape an anisotropy of the Fermi-surface emerges from our theory based on a single-band $t$-$J$ model due to the interplay between the symmetries of the spin liquid and the internal structure of the fermions with no further degrees of freedom needed, we in this sense provide microscopic support for the conjectured existence of a FL* in doped spin liquids, which may in turn be the origin of the pseudogap phase observed in the cuprates~\cite{senthil2003,yang2006,kaul2007,qi2010,moon2011,punk2012, vojta2012,punk2015,Zhang2020,bonetti2024}. We additionally expect that the Fermi surface of a FL$^{*}$ is highly anisotropy, unless the pocket is located at a high-symmetry point, because the intensity of the ARPES signal is determined by this subtle interplay. 

In particular, our theory 
gives a precise  mechanism for the binding of holons and spinons into new fermions, which forms the basis for a number phenomenological theories such as quantum dimer models~\cite{qi2010,vojta2012,punk2015}. 
It would in this connection be interesting to explore if and how the emerging fermions in our theory are connected to the "ancilla" model,
where electrons are coupled to qubits in two layers added as a mathematical tool to model strong electron correlations~\cite{Zhang2020,bonetti2024, christos2024}. Quantum gas microscopy with optical lattices makes it possible to probe the real-space properties of electronlike quasiparticles in the $t$-$J$-model via the spin-hole correlation function, as has already been done for antiferromagnets~\cite{koepsell2019a,ji2021a,koepsell2021a,Lebrat2024,Prichard2024}. Here, it should be noted that the temperatures required to reach the pseudogap phase are significantly higher than those required to achieve superconductivity. 

Our theoretical framework  opens up new ways to explore hole motion in quantum spin liquids and its possible 
relation to pseudogap physics. It is, therefore, interesting to explore the properties of the emerging 
fermions for different values of the parameters entering the  $t$-$J$ model.
Also, even though it is often argued that a 
single-band model captures the essential physics of the cuprates, they are clearly very complex, and 
an important future research direction is, therefore, to 
extend  the $t$-$J$ model to
include, for example, several bands. 
Exploring possible Cooper pairing instabilities of the 
emerging fermions and whether they occur in the $d$-wave channel would provide strong 
indications that the FL* is indeed a parent state of high-$T_c$ superconductors. We note that recent density-matrix renormalization group (DMRG) calculations strongly indicate an instability toward pairing for the same extended $t$-$J$ model as considered here within the same parameter range, using cylinders of up to six-site circumference~\cite{jiang2021a}. A pronounced asymmetry between electron-doped and hole-doped systems was predicted in these calculations, and valuable insights would, therefore, be obtained by performing a similar analysis within our framework with a positive 
$t_2/t_1 = +\sqrt{J_2/J_1}$  corresponding to electron doping. 
Another question concerns if our theoretical framework predicts a quantum phase transitions from the FL* to a conventional 
Fermi liquid at a critical hole concentration as observed experimentally~\cite{annurev:/content/journals/10.1146/annurev-conmatphys-031218-013210}. One should also explore if 
the lifetime and in-plane conductivity of the emerging fermions exhibit Fermi-liquid-like 
behavior as has been observed  in optical conductivity and 
magnetoresistance measurements~\cite{doi:10.1073/pnas.1218846110,doi:10.1073/pnas.1602709113}.

Since the pseudogap phase seems to be a universal phenomenon appearing in many different materials, a key problem concerns whether the emerging fermions
are a robust feature of mobile holes in QSLs. More broadly, it is interesting to investigate whether such fermions exist in other kinds of 
lattice geometries and QSLs. This naturally raises the question of the effects  of gauge fluctuations. While they likely are not important for the 
$\mathbb{Z}_2$ QSL studied here due to their gapped nature making  the low-energy spectrum of the mean-field description similar to that found in variational Monte Carlo studies~\cite{ferrari2019}, 
gauge fluctuations 
 may be important for, e.g.\  U$(1)$ QSLs in triangular lattices~\cite{wen2002, drescher2022, iqbal2016}.   With this in mind it would be very useful to compare results from our theory with those from numerical methods such as matrix-product-state-based dynamical algorithms~\cite{kadow2022,kadow2024} or the variational Monte Carlo method~\cite{charlebois2020}. Additionally, as spin liquid phases become unstable with increasing doping \cite{PhysRevB.86.085145, PhysRevLett.127.187003, PhysRevB.84.174409}, it would be  informative to determine the  pairing fields $\Delta_{\bk}$ as a function of doping, since the bound state formation and spectral properties are dependent on these; see Appendix \ref{app.LS}.
Finally, DMRG studies suggest the use of a broad spectral response arising from the fractionalization of electrons into uncorrelated holons and spinons as a signature of a QSL~\cite{laeuchli2004, kadow2022, kadow2024}. Our results, on the other hand, show that this is not always the case, and one should, therefore, explore how common the emerging fermions are.

\begin{acknowledgments}
J.\ H.\ N.\ and G.\ M.\ B.\ acknowledge support by the Novo Nordisk Foundation Project (Grant No. 0086599). K.K.N. acknowledges support by the Carlsberg Foundation through a Reintegration Fellowship (Grant No. CF24-1214). L.B. was supported by the NSF Condensed Matter and Materials Theory program under Grant No. DMR-2419871, and by the Simons Collaboration on Ultra-Quantum Matter, through a grant from the Simons Foundation (Grant No. 651440). J.H.N.
thanks Jens Paaske for useful comments and discussions.
\end{acknowledgments}

\clearpage


\appendix

\section{Open system dynamics} \label{app.NH}
In this appendix we further examine the consequences of considering the spinon and holon to be interacting in an open system where the holon quasiparticle can decay. To understand and perform calculations for this open system, we follow the approach described in Ref. \cite{brody2014}. We compare these results with calculations where the decay of the holon quasiparticle is omitted.

\begin{figure}[t!]
	\begin{center}
	\includegraphics[width=0.49\textwidth]{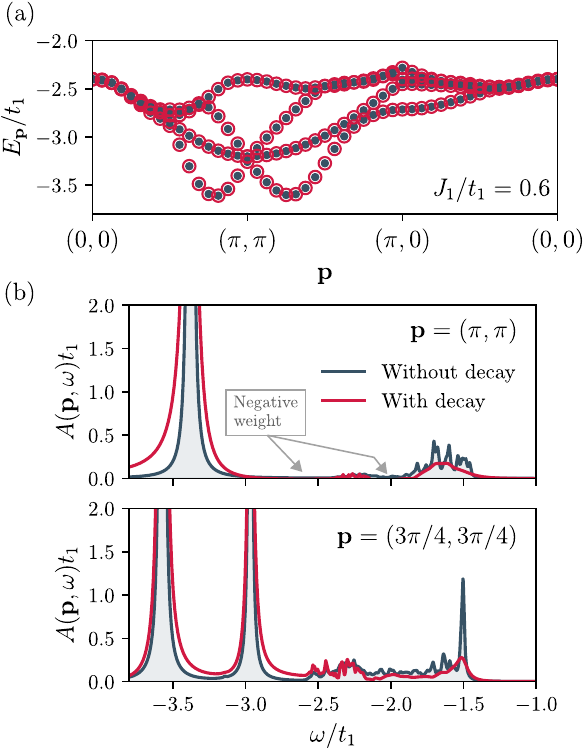}
	\end{center}
	\caption{ (a) The four lowest energies found by solving the effective Hamilton Eq. \eqref{Schrodinger} with decay (red circles) and without decay (blue balls). Calculated for a system size of 32x32, (b) shows the spectral function calculated with Eq. \eqref{eq.spec_open} in red and calculated with Eq. \ref{eq.spec} in blue for the indicated momenta. Gray arrows indicate areas of negative spectral weight when decay is considered.}
	\label{Fig.compareSpec}
\end{figure}

Solving the effective Schr\"odinger, we retrieve complex energies where the real part reflects the observable energy of the state and the imaginary part reflects decay. In Fig. \ref{Fig.compareSpec}(a), we compare the bottom of the energy spectrum when considering decay (blue balls) and when decay is ignored (red circles). This shows no visible difference, and we find the average difference between the two ways of calculating the energies to be $\sim10^{-4}\cdot t_{1}$.

Including the decay of the holons means that the decay of the spinon-holon states is given by the imaginary part of the energy. Using the framework developed in Ref. \cite{brody2014}, we define the right and left states as $\hat{H}\ket{\Psi^{n}_{\bp,r}} = E^{n}_{\bp}\ket{\Psi^{n}_{\bp,r}}$ and $\bra{\Psi^{n}_{\bp,l}}\hat{H} = E^{n}_{\bp}\bra{\Psi^{n}_{\bp,l}}$ respectively. This leads to 
\begin{align}
    \ket{\Psi^{n}_{\bp,r}} &= \sum_{\bk}\phi^{n}_{\bp,\bk}\tilde{b}^{\dagger}_{\bp-\bk}\hat{\gamma}^{\dagger}_{\bk,\downarrow}\ket{\text{QSL}} \\ \nonumber
    \ket{\Psi^{n}_{\bp,l}} &= \sum_{\bk}[\phi^{n}_{\bp,\bk}]^{*}\tilde{b}^{\dagger}_{\bp-\bk}\hat{\gamma}^{\dagger}_{\bk,\downarrow}\ket{\text{QSL}}
\end{align}
and $\sum_{\bp,n} \ket{\Psi^{n}_{\bp,r}}\bra{\Psi^{n}_{\bp,l}} = \mathds{1}$. With use of the Lehmann representation of the Green's function the spectral function reads
\begin{align}
    A(\bp,\omega) = - \frac{2}{N}\sum_{n}{\rm Im}\Bigg[\frac{1}{\omega-E_{\bp}^{n}} \left(\sum_{\bk}v_{\bk}\phi^{n}_{\bp,\bk}\right)^{2}\Bigg],
    \label{eq.spec_open} 
\end{align}
 when decay is considered. In Fig. \ref{Fig.compareSpec}(b), we see the spectral function calculated with Eq. \eqref{eq.spec_open} in red. In blue, we see the spectral function calculated with Eq. \eqref{eq.spec}. This again shows that including the decay does not change the behavior at the bottom of the spectrum. Considering decay, we also see that the spectral function can take negative values in the scattering continuum. This is not uncommon when considering a non-Hermitian Hamiltonian and illustrates that these states consist of a superposition that includes holon states with a large decay. 

\begin{figure}[t!]
	\begin{center}
	\includegraphics[width=0.49\textwidth]{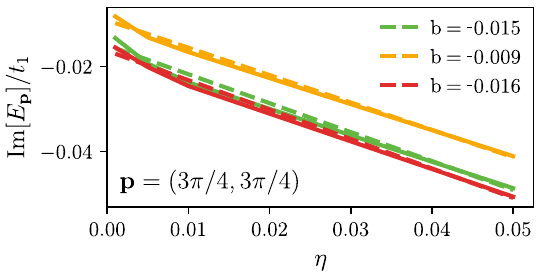}
	\end{center}
	\caption{As a function of the artificial broadening $\eta$, we plot the imaginary part of the energy for the three bound states. With dashed lines, we plot the linear fit $a\cdot\eta + b$. The color code is the same as used in Fig. \ref{Fig.EnergySpectrum}.}
	\label{Fig.eta}
\end{figure}

We also find that the bound states are broadening by including decay, which comes from the states having a finite overlap with holon states with a finite decay. In Fig. \ref{Fig.eta}, we investigate the decay of the bound state, Im$[E^{n}_{\bp}]$, for $\bp=(3\pi/4,3\pi/4)$ as a function of the infinitesimal broadening, $\eta$. Here, we see that it converges to a finite value as $\eta\rightarrow0$. This value is $\sim10^{-2}\cdot t_{1}$, which for the ground state is a factor of $10^{2}$ larger than the energy gap to the scattering continuum making this quasiparticle state well-defined.

\section{Iterative approach to calculating the Green's function of the hole} \label{app.FullCalc}

In this appendix we calculate the Green's function of the hole, Eq. \eqref{eq.prop_hole_rot}, iteratively by setting up a Bethe-Salpeter equation for the scattering matrix. We first expand the Green's function in terms of Feynman diagrams as 
\begin{align}
	\includegraphics[width=0.42\textwidth]{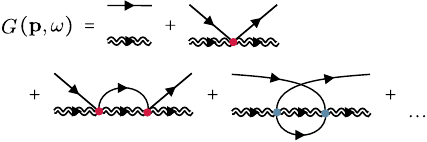}
    \label{eq.FeynProp}
\end{align}
with the red vertex corresponding to the $h$ vertex and the blue vertex corresponding to the $g$ vertex in Eq. \eqref{Eq.H_inter}. With $J_{1}/J_{2}=t_{1}^{2}/t_{2}^{2}=0.5$ all the vertices become proportional to $t_{1}/J_{1}$ leaving no small parameter to expand in. We initiate these calculations by calculating the holon's Green's function, the full wavy line in Eq. \eqref{eq.FeynProp}, as done in Ref. \cite{nyhegn2025}. Wanting to capture nonperturbative results, we then set up a Bethe-Salpeter equation for the scattering matrix defined in Fig. \eqref{eq.GreenScat}(c). This means we include a subset of all diagrams, but we include these in a self-consistent way, and in such a way that we retrieve a conserving approximation according to Ref. \cite{baym1961,baym1962}. Now, by calculating the scattering matrix $\Gamma$ iteratively until convergence, we can calculate the Green's function by inserting this into the expression in Fig. \eqref{eq.GreenScat}(a). 

\begin{figure}[t!]
	\begin{center}
	\includegraphics[width=0.49\textwidth]{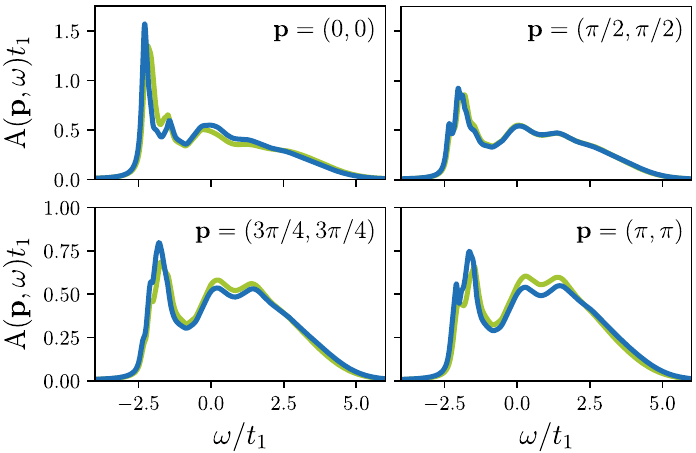}
	\end{center}
	\caption{For a system size of $8\times8$, in blue we plot the hole spectral function obtained by use of Eq. \eqref{eq.T2} as the scattering matrix. Here, $J_{1}/t_{1}=0.6$, $\eta=0.1$, and the crystal momenta are indicated in the respective panels. In green, we plot the spectral function when only considering the co-propagating term. }
	\label{Fig.T2}
\end{figure}

Before performing the full calculations, let us first understand the impact of the two different types of diagram considered. To do this, we separate the scattering matrix into 
\begin{align}
	\includegraphics[width=0.45\textwidth]{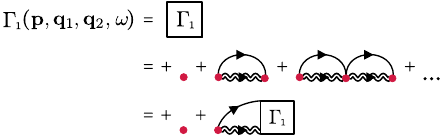}
    \label{eq.T1}
\end{align}
and
\begin{align}
	\includegraphics[width=0.45\textwidth]{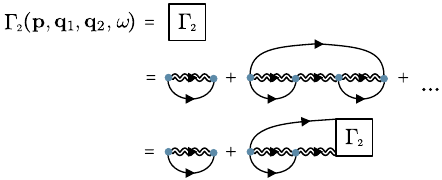}
    \label{eq.T2}
\end{align}
which leaves out the diagrams mixing the two types of diagram, but allows an investigation of the individual contributions. By setting $\Gamma = \Gamma_{2}$ in the expression for the Green's function, we find for $J_{1}/t_{1}=0.6$ the results presented in Fig. \ref{Fig.T2}. In green, we also present the first term in Eq. \eqref{eq.FeynProp}, which is the noninteracting co-propagating term. Now, while these calculations converge and fulfill the sum rule with deviations of less than 1\%, we see that including the interactions barely changes the spectral function. This indicates that most of the dynamics is governed by the co-propagating term, meaning that the holon and the spinon do not bind and that spin-charge separation is a good assumption. Similar results were found for different interaction strengths and crystal momenta.

\begin{figure}[t!]
	\begin{center}
	\includegraphics[width=0.49\textwidth]{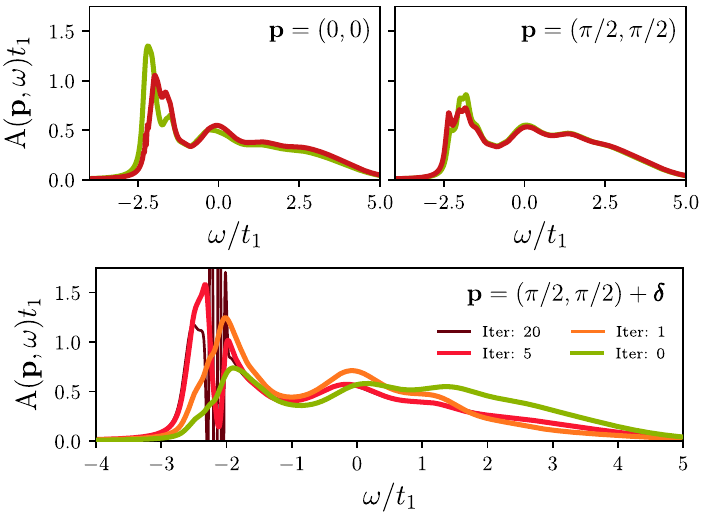}
	\end{center}
	\caption{In red, we plot the hole spectral function computed with Eq. \eqref{eq.T2} as the scattering matrix. For comparison with the results from Fig. \ref{Fig.T2}, we use a system size of $8\times 8$ for (a) and (b) and use a system size of $16\times16$ for (c). $J_{1}/t_{1}=0.6$, $\eta=0.1$, and the crystal momenta are indicated in the respective panels with $\boldsymbol{\delta} = (2\pi/L,2\pi/L)$. In green, we plot the spectral function when considering only the co-propagating term. The calculation does not converge. In panel (c) we, therefore, plot several calculations terminated after a different number of iterations (Iter).}
	\label{Fig.T1}
\end{figure}

\begin{figure}[t!]
	\begin{center}
	\includegraphics[width=0.48\textwidth]{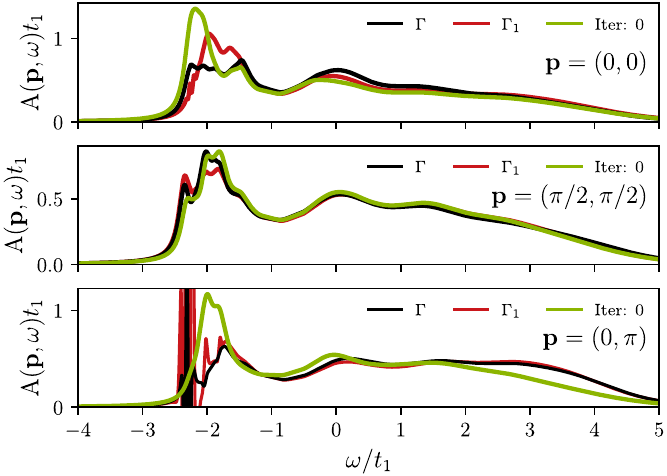}
	\end{center}
	\caption{Spectral function calculated with the full scattering matrix from Fig. \eqref{eq.GreenScat}(c) (black), with Eq. \eqref{eq.T1} (red), and with the co-propagating term (green). These calculations are performed for a system size of $8\times8$ and for $J_{1}/t_{1}=0.6$, $\eta=0.1$, and the indicated momenta. }
	\label{Fig.Tfull}
\end{figure}

Instead, choosing $\Gamma = \Gamma_{1}$, we obtain the spectral function presented in Fig. \ref{Fig.T1}. For the plots at the top, we again find that the co-propagating term dominates and that the interactions do not change much. In contrast, the lower panel shows that the interactions move spectral weight to the bottom of the spectrum, but ultimately diverge. For each additional iteration, the frequency and amplitude of the oscillations increase.  We find similar divergences for crystal momenta in the vicinity of $\bp = (\pi,\pi)$. Finally, including the full scattering matrix from Fig. \eqref{eq.GreenScat}(c), we see in Fig. \ref{Fig.Tfull} that the scattering matrix from Eq. \eqref{eq.T1} captures the same divergences, and hence the essential dynamics of these nonperturbative results.

The conclusion from this analysis is that away from $\bp=(\pi,\pi)$, the Green's function is well-approximated by the co-propagating term. Close to $\bp=(\pi,\pi)$, we see that the type of interactions considered in Eq. \eqref{eq.T2} is irrelevant, and instead, the scattering interactions considered in Eq. \eqref{eq.T2} are relevant. By considering these, we find indications of nonperturbative features occurring around $\bp=(\pi,\pi)$. These features come in the form of divergences, and we believe these stem from the approach trying to perturbatively capture a pole, hence a spinon-holon bound state. 
Additional support for this conclusion is found in the Sec. \ref{sec.BoundState} of the main text, where the analysis predicts bound states at the crystal momenta where the divergences occur.


\section{Length of singlets} \label{app.LS}

One of the motivations for using a BCS-type Hamilton to describe the resonating valence bond (RVB) state is that when the Gutzwiller operator is applied on the BCS-wave function, the Cooper pairs become resonating singlets and we obtain an RVB state \cite{baskaran1987}. The pairing potential, therefore, determines the nature of the singlet pairings. In this appendix, we explore the results of having a RVB background with longer singlets. We do this by decreasing $\Delta_{1}/\epsilon$ while keeping $\Delta_{1}/\Delta_{2}$ constant in Eq. \eqref{Eq.spinon}. In doing this, we soften the spinon spectrum and change the average length of the singlets as described in Ref. \cite{nyhegn2025}. 
\begin{figure}[t!]
	\begin{center}
	\includegraphics[width=0.48\textwidth]{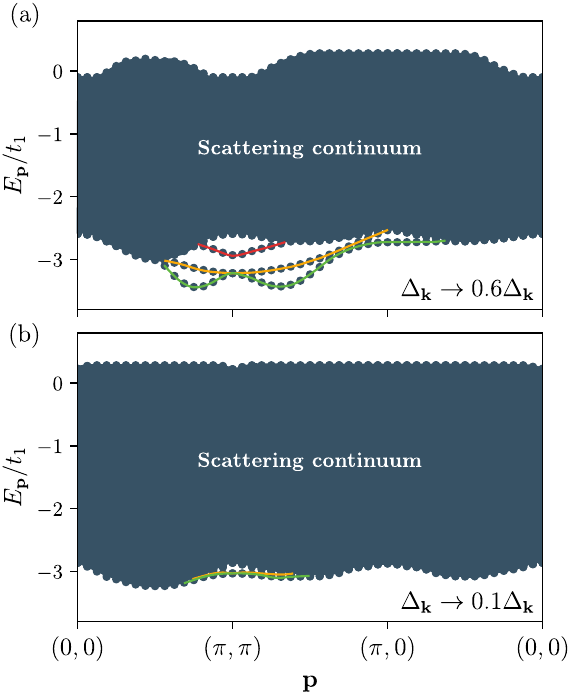}
	\end{center}
	\caption{With $J_{1}/t_{1}=0.6$ and $\Delta_{1}/\Delta_{2}$ constant, the energy spectrum for $\Delta_{1}/\epsilon=1.08$ and $\Delta_{1}/\epsilon=0.18$ is shown in (a) and (b) respectively. The text in the panel states the overall scaling of the pairing potential. The calculations are performed for a system size of $32\times32$. Red, yellow, and green lines highlight bound states below the scattering continuum. }
	\label{Fig.EnergySpectrumScale}
\end{figure}

In Fig. \ref{Fig.EnergySpectrumScale}(a) we see the energy spectrum calculated with $\Delta_{1}/\epsilon=1.08$, which increases the average length of the singlets by $\sim1.5$ compared with $\Delta_{1}/\epsilon=1.8$ used for the calculations in Sec. \ref{sec.BoundState} and \ref{sec.observables} of the main text. Compared with the results from Fig. \ref{Fig.EnergySpectrum}(a),  Fig. \ref{Fig.EnergySpectrumScale}(a) shows that this environment decreases the binding energy. In Fig. \ref{Fig.EnergySpectrumScale}(b), we have further decreased it to $\Delta_{1}/\epsilon=0.18$, and here we see that the red band completely disappears into the scattering continuum.

\begin{figure}[t!]
	\begin{center}
	\includegraphics[width=0.48\textwidth]{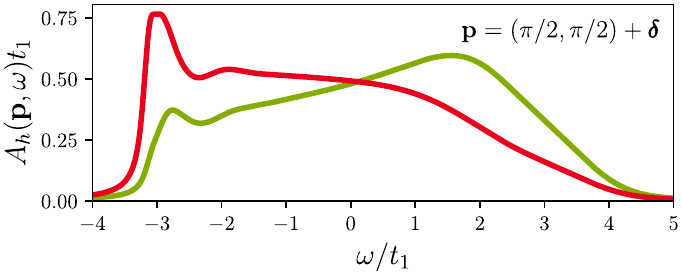}
	\end{center}
	\caption{In red, we plot the spectral function calculated with the iterative method described in Appendix \ref{app.FullCalc}. We performed 50 iterations and found convergence. The co-propagating term is plotted in green, and the calculations are performed with $\Delta_{1}/t_{1}=0.18$, $J_{1}/t_{1}=0.6$, and $\boldsymbol{\delta} = (2\pi/L,2\pi/L)$, for a system size of $16\times16$.}
	\label{Fig.BS_scale}
\end{figure}

To further establish these results, we show in Fig. \ref{Fig.BS_scale} the results from the same calculations as those in Fig. \ref{Fig.T1}(c) but with $\Delta_{1}/\epsilon=0.18$. While the results in Fig. \ref{Fig.T1} show indications of bound state formation, these do not, and we find a general trend that decreasing $\Delta_{1}/\epsilon$ results in the iterative calculations converging for more crystal momenta in the Brillouin zone. Altogether, this indicates that having a stiffer spinon spectrum with shorter singlets assists the formation of spinon-holon bound states, while systems with longer singlets promote separation of the spinon and the holon, yielding a broad spectral response.

\section{Gauge Averaging} \label{app.gaugeAVG} 

In Sec. \ref{sec.BoundState} and \ref{sec.observables} of the main text, we perform a gauge averaging to recover the full square lattice symmetry of all observables.  Here, we elaborate on this procedure and where the symmetry breaking originates.

As explained in the Sec. \ref{sec.model}, the optimal mean-field description of the $\mathbb{Z}_2$ QSL is
\begin{align}
	\hat{H}^\text{mf}_{J} =  \sum_{\bk, \sigma} \epsilon_{\bk} \hat{ f }^{\dagger}_{\bk,\sigma} \hat{ f }_{\bk,\sigma}  + \sum_{\bk} \left( \Delta_{\bk} \hat{f}^{\dagger}_{\bk,\uparrow} \hat{f}^{\dagger}_{-\bk,\downarrow} + \text{h.c.}\right).
	\label{app.spinon} 
\end{align}
When projected onto the half-filled subspace, the Cooper pairs of the BCS wave function turn into the singlets covering the lattice.   The presence of both $d_{x^2-y^2}$ and $d_{xy}$ pairing apparently breaks reflection symmetry; for example, $x\rightarrow -x$ changes the relative sign of the two pairing components (as does $x\leftrightarrow y$).  At half-filling, however, it is known that gauge-invariant observables are fully symmetric, as discussed in Ref.\cite{wen2002,hu2013}.  In particular, the relative sign of the two pairing terms is modified by the pure gauge transformation
\begin{equation}
    \hat{f}_{{\bf i}\alpha} \rightarrow \epsilon_{\alpha\beta} \hat{f}_{{\bf i}\beta}^\dagger (-1)^{|\bf i|}.
    \label{eq:su2gauge}
\end{equation}
This is part of the SU(2) gauge symmetry of the spin-liquid ansatz $\hat{{\bf S}}_{\bf i} = \frac12 \hat{f}_{\bf i}^\dagger \bm{\sigma} \hat{f}_{\bf i}$, so at half-filling, for which all observables can be expressed in terms of spin operators, the parity noninvariance of $\hat{H}^\text{mf}_J$ is purely a gauge artifact.

However, in the larger Hilbert space of the $t$-$J$ model, and with use of the U(1) slave boson ansatz $\hat{c}_{{\bf i},\alpha} = \hat{b}_{\bf i}^\dagger \hat{f}_{{\bf i},\alpha}$, the transformation in Eq.~\ref{eq:su2gauge} is no longer possible (because there is no corresponding transformation of the holon operators that leaves the electron invariant).  Therefore, our theory describes a situation in which the parity symmetry becomes truly broken for physical observables.  

Because the underlying $t$-$J$ model has parity symmetry, the parity breaking must be spontaneous, and hence in repeated experiments the choice of the parity-broken state will vary from realization to realization. To account for this, we perform a gauge averaging over the different gauges of $\Delta_{\bk}$. Specifically, we perform two calculations -- one using $\Delta_{(k_{x},k_{y})}$ and the other using $\Delta_{(k_{y},k_{x})}$ -- and then we take the average of the two results. For the relative wave function $\phi^{n}_{\bp,\br}$ in Fig. \ref{Fig.EnergySpectrum}(c), this means we compute $|\phi^{n}_{\bp,\br}|^{2}$ for both gauges individually, and then we take the average.

In principle, it is possible to choose a different formulation of the doped system which retains the full symmetry of the half-filled spin-liquid state, but this requires going beyond the U(1) slave boson theory to the SU(2) slave boson theory\cite{wen1996theory}, in which there are two flavors of holon operators.  We leave such an approach to future work.

\clearpage

\bibliography{SpinLiquid}


\end{document}